\documentclass[structabstract]{aa}  
\usepackage{graphicx}
\usepackage{amsmath}
\usepackage{natbib}
\bibpunct{(}{)}{;}{a}{}{,}
\usepackage{txfonts}
\begin{document}
   \title{Global dynamo models from direct numerical simulations and 
          their mean-field counterparts}

   \author{M. Schrinner\inst{1}}

   \institute{MAG (ENS / IPGP), LRA, Ecole Normale Sup\'erieure, 24 rue Lhomond,
              75252 Paris Cedex 05, France\\
              \email{martin@schrinner.eu}
             }

   \date{Received; accepted}

 
  \abstract
   {The recently developed test-field method permits to compute dynamo 
    coefficients from global, direct numerical simulations. The 
    subsequent use of these parameters in mean-field models enables us to 
    compare self-consistent dynamo models with their mean-field 
    counterparts. So far, this has been done for a simulation of rotating
    magnetoconvection and a simple benchmark dynamo, which are both
    (quasi-)stationary.}  
   {It is shown that chaotically time-dependent dynamos in a low Rossby number 
    regime may be appropriately described by corresponding 
    mean-field results. Also, 
    it is pointed out under which conditions mean-field models do not match 
    direct numerical simulations.}
   {We solve the equations of magnetohydrodynamics (MHD) in a rotating, 
    spherical shell in the Boussinesq approximation. Based on this, we compute 
    mean-field coefficients for several models with the help of the previously 
    developed test-field method. The parameterization of the mean electromotive
    force by these coefficients is tested against direct numerical simulations.
    In addition, we use the determined dynamo coefficients in mean-field models
    and compare the outcome with azimuthally averaged fields from direct 
    numerical simulations.}
   {The azimuthally and time averaged electromotive force in fast rotating 
    dynamos is sufficiently well parameterized by the set of 
    determined mean-field coefficients. In comparison to the previously 
    considered (quasi-)stationary dynamo, the chaotic time-dependence leads 
    to an improved scale separation and thus to a better agreement between 
    direct numerical simulations and mean-field results.}
   {}


   \maketitle
%
%
%
%
\section{Introduction}
 Mean-field electrodynamics provides conceptual understanding of dynamo
 processes generating coherent, large-scale magnetic fields in planets, 
 stars and galaxies \citep{raedler80,moffatt}. 
 But, it suffers from a number of free parameters,  
 which are only poorly known under astrophysically relevant conditions. 
 We briefly summarize the essentials of mean-field theory in order to 
 specify this point. Within a mean-field approach, the velocity and the 
 magnetic field are usually split in large-scale mean fields, 
\(\overline{\vec{V}},\overline{\vec{B}}\), and residual  
 parts, \(\vec{v},\vec{b}\), varying on much shorter length or time 
 scales, 
 \begin{equation}
   \vec{V}=\overline{\vec{V}}+\vec{v},\quad\vec{B}=\overline{\vec{B}}+\vec{b}.
   \label{eq2}
 \end{equation}
 Then, the evolution of the mean field is governed by 
 \begin{equation}
   \frac{\partial{\overline{\vec{B}}}}{\partial t}
   =\nabla\times(\vec{\mathcal{E}}+\overline{\vec{V}}\times\overline{\vec{B}}
   -\eta\nabla\times\overline{\vec{B}}).
   \label{eq4}
 \end{equation}  
 In the above equation, the mean electromotive force is defined as 
 \(\vec{\mathcal{E}}=\overline{\vec{v}\times\vec{b}}\) and \(\eta\) is
 the magnetic diffusivity. Equation (\ref{eq4}) is
 coupled with an equation for the residual field
 \begin{equation}
   \frac{\partial\vec{b}}{\partial t}
   =\nabla\times(\vec{G}+\overline{\vec{V}}\times\vec{b}
   +\vec{v}\times\overline{\vec{B}}
   -\eta\nabla\times\vec{b}),
   \label{eq6}
 \end{equation}
in which \(\vec{G}\) is defined as \(\vec{G}=\vec{v}\times\vec{b}-\overline{\vec{v}\times\vec{b}}\).
The above system of equations is equivalent to the usual induction equation 
and no approximation has been applied so far. An important simplification of 
the mean-field approach consists in considering equation (\ref{eq4}) only,
which requires to express \(\vec{\mathcal{E}}\) as a linear functional 
of \(\vec{v},\overline{\vec{V}}\) and \(\overline{\vec{B}}\). This is 
justified with regard to (\ref{eq6}), and we may write
\begin{equation}
\vec{\mathcal{E}}_i=\vec{\mathcal{E}}_i^{(0)}+\int\int
\tens{K}_{ij}(\vec{x},\vec{x}',t,t')\,\overline{\vec{B}}_j(\vec{x}',t')
\,d^3x'\,dt',
\label{eq8}
\end{equation}
in which \(\tens{K}\) denotes some integral kernel and we refer to 
cartesian coordinates for the moment. 
The term \(\vec{\mathcal{E}}^{(0)}\) 
in (\ref{eq8}) accounts for small-scale dynamo action and is not considered 
here; it will be discussed later. 
In addition, we assume that \(\vec{\mathcal{E}}\) depends only  
instantaneously and nearly locally on \(\overline{\vec{B}}\). This is a 
crucial assumption, it will be referred to as the assumption of scale 
separation in the following. Therefore, \(\overline{\vec{B}}\) 
in (\ref{eq8}) may be replaced
by its rapidly converging Taylor series expansion at \(\vec{x}\),
\begin{equation}
  \overline{\vec{B}}_j(\vec{x}',t)=\overline{\vec{B}}_j(\vec{x},t)
  +(x'_k-x_k)\frac{\partial\overline{\vec{B}}_j(\vec{x},t)}
  {\partial x_k}+\cdots\,,
\label{eq10}
\end{equation}
and taken out of the integral:
\begin{equation}
  \vec{\mathcal{E}}_i=\tens{a}_{ij}\overline{\vec{B}}_j+\tens{b}_{ijk}\frac{\partial \overline{\vec{B}}_j}{\partial x_k}+\cdots\,,
  \label{eq12}
\end{equation}
with
\begin{eqnarray}
  \tens{a}_{ij} & = & \int\int
  \tens{K}_{ij}(\vec{x},\vec{x'},t,t')\,d^3x'dt'\,, \label{eq14}\\
\tens{b}_{ijk} & = & \int\int
\tens{K}_{ij}(\vec{x},\vec{x'},t,t') (x_k'-x_k)\,d^3x'dt'
\label{eq16}
\end{eqnarray}
\begin{equation*}
\cdots
\end{equation*}
The dynamo coefficients \(\tens{a}\) and \(\tens{b}\) 
defined in (\ref{eq14}) and (\ref{eq16}) together with the expansion 
(\ref{eq12}) may finally be used to solve equation (\ref{eq4}). Moreover, 
\(\tens{a}\) and \(\tens{b}\) are directly linked to dynamo processes, 
i.e. to the generation, advection and 
diffusion of the mean magnetic field, and thus provide physical concepts to 
explain dynamo action \citep[e.g.][]{raedler95}. 
Unfortunately, they are not known in general and previous work 
often relies on arbitrary assumptions on them.    

The test-field method by \cite{schrinner05,schrinner07} permits to compute 
\(\tens{a}\) and \(\tens{b}\) from direct numerical 
simulations. It was first applied to a simulation of rotating 
magnetoconvection \citep{olson99}  and a simple geodynamo simulation 
\citep{christensen01}, which are both stationary except for 
an azimuthal drift. In the past three years, the test-field method 
was intensively used to compute mean-field coefficients for 
box simulations \citep[e.g.][]{sur08,gressel08,branden09,kaepy09,raedler09},
which are not in the focus of this paper. 
Although the test-field method proved to be reliable, 
the parameterization of the electromotive force for the steady geodynamo 
model was not satisfactory. The expansion in (\ref{eq12}) does not converge 
for this steady example because of a non-sufficient scale separation
\citep{schrinner07}.  
In this paper, we revisit the problem for chaotically time-dependent dynamos. 
We test azimuthally and time averaged electromotive-force vectors against their
parameterization based on corresponding dynamo coefficients and compare 
azimuthally and time-averaged fields from direct numerical simulations with 
mean-field results.   
\section{Dynamo simulations}
We solve the equations of magnetohydrodynamics (MHD) in the Boussinesq 
approximation for a conducting fluid in a rotating spherical shell as given by
\cite{olson99},
\begin{align}
E\left(\frac{\partial\vec{V}}{\partial t}+\vec{V}\cdot\nabla\vec{V}-\nabla^2\vec{V}\right)
+2\vec{z}\times\vec{V}+\nabla P =\nonumber\\
Ra\frac{\boldsymbol{r}}{r_o}T +\frac{1}{Pm}(\nabla\times\vec{B})\times\vec{B}\label{eq18}\\
\frac{\partial\vec{B}}{\partial t}= \nabla\times(\vec{v}\times\vec{B})
+\frac{1}{Pm}\nabla^2\vec{B}\label{eq20}\\
\frac{\partial T}{\partial t}+\vec{v}\cdot\nabla T  =
\frac{1}{Pr}\nabla^2 T.\label{eq22}
\end{align} 
The coupled differential equations for the the velocity \(\vec{V}\), the 
magnetic field \(\vec{B}\), and the temperature \(T\) are governed by four
parameters. These are the Ekman number \(E=\nu/\Omega L^2\), 
the (modified) Rayleigh number \(Ra=\alpha_T g_0\Delta T L/\nu\Omega\), 
the Prandtl number \(Pr=\nu/\kappa\) and the magnetic Prandtl number 
\(Pm=\nu/\eta\). In these expressions,  \(\nu\) denotes the kinematic 
viscosity, \(\Omega\) the rotation rate, \(L\) the shell width, 
\(\alpha_T\) the thermal expansion coefficient, \(g_0\) is the gravitational 
acceleration at the outer boundary, \(\Delta T\) stands for the temperature 
difference between the spherical boundaries, \(\kappa\) is the 
thermal and \(\eta=1/\mu\sigma\) the magnetic diffusivity with the magnetic 
permeability \(\mu\) and the electrical conductivity \(\sigma\). Besides these
input parameters, we introduce the magnetic Reynolds number 
\(Rm=V_\mathrm{rms}\,L/\eta\), and the local Rossby number 
\(Ro_l=V_\mathrm{rms}/(\Omega L)\cdot(\overline{l}/\pi)\) as important output 
parameters. In the latter expressions, \(V_\mathrm{rms}\) denotes the 
rms-velocity and \(\pi/\overline{l}\) is the mean half-wavelength of the flow 
derived from the kinetic energy spectrum \citep{christensen06}.   

No-slip mechanical boundary conditions were chosen for all simulations
presented here and the magnetic field continues as a potential field outside 
the fluid shell. Convection is driven by an imposed temperature difference 
between the inner and the outer shell shell-boundary.     
\section{Computation of dynamo coefficients}
\label{sec:6}
In order to determine the dynamo coefficients \(\tens{a}\) and \(\tens{b}\), 
we apply the test-field method explained in full detail in \cite{schrinner07}. 
The principal idea of this approach is to measure the mean electromotive force 
generated by the interaction of the flow with an arbitrarily imposed test 
field, \(\overline{B}_\mathrm{T}\). The imposed test field appears as an 
inhomogeneity in equation (\ref{eq6}), 
\begin{equation}
  \frac{\partial\vec{b}}{\partial t}
  -\nabla\times(\vec{G}+\overline{\vec{V}}\times\vec{b}
  -\eta\nabla\times\vec{b})
  =\nabla\times\vec{v}\times\overline{\vec{B}}_\mathrm{T},
  \label{eq24}
\end{equation}
which is solved simultaneously with (\ref{eq18}-\ref{eq22}). The 
electromotive force due to a given \(\overline{\vec{B}}_T\) may then be 
computed as \(\vec{\mathcal{E}}_\mathrm{T}=\overline{\vec{v}\times\vec{b}}\)
with \(\vec{b}\) resulting from (\ref{eq24}). Finally, 
\(\vec{\mathcal{E}}_T\) is used to solve relation (\ref{eq12}) for the dynamo 
coefficients. In order to close the resulting system of linear equations and 
to determine all components of  \(\tens{a}\) and \(\tens{b}\), the numerical 
experiment (\ref{eq24}) must be repeated several times with different  
fields \(\overline{\vec{B}}_T\).  

So far, averaged quantities labelled by an overbar are axisymmetric fields
whereas  \(\vec{v}\) and \(\vec{b}\) in (\ref{eq24}) are non-axisymmetric 
residuals. For a stochastically stationary but nevertheless 
chaotically time-dependent flow, \(\vec{\mathcal{E}}\) and also 
the mean-field coefficients vary in time. Thus, we introduce an additional
time averaging indicated by brackets, \(<\cdots>\), and write
\begin{equation}
\vec{\mathcal{E}}\approx\tens{a}<\!\overline{\vec{B}}\!>+\tens{b}<\!\nabla\overline{\vec{B}}\!>+\cdots
\label{eq26}
\end{equation}
instead of (\ref{eq12}). In the above equation, \(\vec{\mathcal{E}}\) 
has been time averaged,
\begin{equation}
\mathcal{E}=<\!\overline{\vec{v}\times\vec{b}}\!>,
\label{eq30}
\end{equation}
and also \(\tens{a}\) and \(\tens{b}\) denote time averaged tensors. 
Furthermore, we interpret averages in (\ref{eq4}) as combined azimuthal 
and time averages. Both is formally not correct. However, the approximations 
applied can be justified, if

i) temporal fluctuations of the axisymmetric velocity and magnetic field 
 are comparatively small, and

ii) time averages of the non-axisymmetric residuals of the velocity and 
    the magnetic field are negligible. 
 
The above assumptions imply that azimuthal and time averaging act in 
a similar way and reinforce each other. As a consequence, mean-field 
coefficients are derived in this context from time averages of vectors 
\(\vec{\mathcal{E}}_\mathrm{T}\) resulting from (\ref{eq24}). We emphasise 
that assumptions i) and ii) are not justified a priori but have to be tested in 
numerical simulations. 
\section{Comparison of mean-field results with direct numerical simulations}
\label{sec:8}
Mean-field coefficients determined from direct numerical simulations may be 
used in mean-field models. Subsequently, azimuthally and time-averaged 
magnetic fields resulting from three-dimensional, self-consistent models may 
be compared with their mean-field counterparts. Mean-field calculations 
presented here are based on equation (\ref{eq4}) written as an eigenvalue 
problem,
\begin{equation}
\sigma\overline{\vec{B}}=\nabla\times\tens{D}\overline{\vec{B}},
\label{eq32}
\end{equation}
in which the linear operator \(\tens{D}\) is defined as
\begin{equation}
\tens{D}\overline{\vec{B}}=\overline{\vec{V}}\times\overline{\vec{B}}+\tens{a}\overline{\vec{B}}+\tens{b}\nabla\overline{\vec{B}}-1/Pm\nabla\times\overline{\vec{B}},
\label{eq34}
\end{equation} 
and \(\tens{a}\) and \(\tens{b}\) are time-averaged dynamo coefficients as 
described above.  The eigenvalues \(\sigma\) lead to a time evolution of each 
mode proportional to \(\exp{(\sigma t)}\). For more details concerning the 
eigenvalue calculation, we refer to \cite{schrinner10b}. 

In addition, the parameterization of the electromotive force by dynamo 
coefficients, i.e. relation (\ref{eq12}), may be tested against the mean 
electromotive force in direct numerical simulations. Again by virtue of 
(\ref{eq26}-\ref{eq30}), we compare \(\vec{\mathcal{E}}\) as given in 
(\ref{eq30}) and
derived directly from numerical models with 
\begin{equation}
\vec{\mathcal{E}}^{(1)}=\tens{a}<\!\overline{\vec{B}}\!>+\tens{b}<\!\nabla\overline{\vec{B}}\!>,
\label{eq36}
\end{equation}
in which \(<\!\overline{\vec{B}}\!>\) and \(<\!\nabla\overline{\vec{B}}\!>\) 
are also obtained from direct numerical simulations. Furthermore, we follow 
\cite{schrinner07} and introduce 
\begin{equation}
\Delta\mathcal{E}=\frac{\int_V\left|\vec{\mathcal{E}}-\vec{\mathcal{E}}^{(1)}\right|\,\mathrm{d}V}{\int_V\left|\vec{\mathcal{E}}\right|\,\mathrm{d}V}
\end{equation}    
in order to quantify errors in the parameterization of \(\vec{\mathcal{E}}\).

\section{Results}
\begin{figure}
  \centering
  \includegraphics{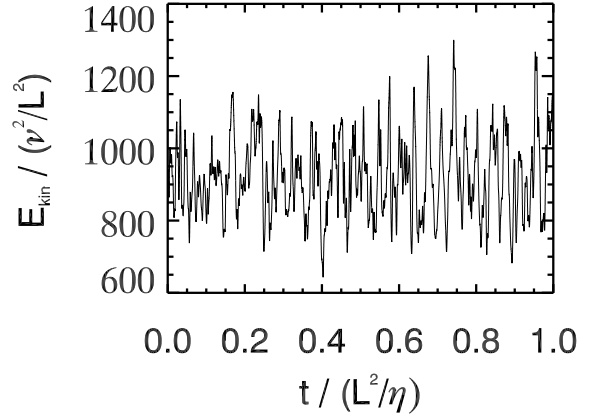}
  \caption{Volume-averaged kinetic energy density versus magnetic diffusion 
   time for model 3.}
  \label{fig7}
\end{figure}

%
\begin{table}
\caption{Overview of the models considered, ordered with respect to their local
Rossby number.}             
\label{table:1}      
\centering          
\begin{tabular}{c c c c c r r r c }     
\hline\hline       
Model & \(E\) & \(Ra\)& \(Pm\)& Mean l&\(Ro_l\)&\(Rm\)&\(\Delta\mathcal{E}\)&\(\lambda/(\eta/L^2)\)\\ 
\hline                    
   1 & \(1\times 10^{-3}\) & 100 & 5 & 5& 0.013& 39& 4.6 &  4.18\\  
   2 & \(1\times 10^{-4}\) & 334 & 3 &  11& 0.014& 123& 1.6 & -7.05\\
   3 & \(1\times 10^{-4}\) & 334 & 2 & 11& 0.015& 86& 1.4 & -3.87\\
   4 & \(3\times 10^{-4}\) & 195 & 3 & 9& 0.019& 67& 1.6 &  2.22\\
   5 & \(3\times 10^{-4}\) & 243 & 2 & 9& 0.024& 56& 1.2 &  0.66\\
   6 & \(3\times 10^{-4}\) & 285 & 2 & 9& 0.026& 61& 1.3 &  0.05\\
   7 & \(3\times 10^{-4}\) & 375 & 1.5& 11& 0.042& 60& 1.1 &  0.78\\
\hline                  
\end{tabular}
\end{table}
Besides the (quasi-)steady benchmark dynamo (model 1), we considered
6 chaotically time-dependent dynamos. The models are 
defined by 4 control parameters; \(E\), \(Ra\), and \(Pm\) were varied 
(see Table \ref{table:1}), whereas the Prandtl number was kept fix and 
set to 1 for all simulations. The local Rossby number is always lower 
than \(0.12\) for the models considered here. They therefore belong to the 
regime of kinematically stable dynamos \citep{schrinner10a}. In order to 
simplify the time averaging, models with low and fairly moderate 
\(Rm\) were chosen. The resulting velocity fields are almost
symmetric with respect to the equatorial plane, except for model 7, and 
convection occurs mainly outside the inner core tangent cylinder. 
Figure \ref{fig7} shows the volume-averaged kinetic energy density for 
model 3 and illustrates the irregular time-dependence of these models.

\begin{figure}
  \centering
  \includegraphics{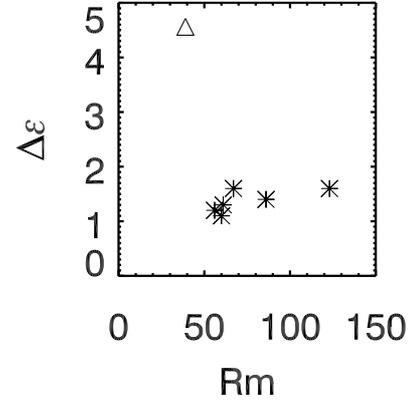}
  \caption{\(\Delta\mathcal{E}\) versus the magnetic Reynolds number 
   for the (quasi)-stationary benchmark dynamo (triangle) and the 6 
   chaotically time-dependent dynamos considered.}
  \label{fig1}
\end{figure}

The parameterization of the mean electromotive force given by (\ref{eq36}) 
improved a lot for the chaotically time-dependent dynamos in 
comparison to the benchmark dynamo; \(\Delta\mathcal{E}\) decreases by more 
than a factor of 4 from model 1 to model 7. The considerable drop 
of \(\Delta\mathcal{E}\) from the steady to the time-dependent models
is most clearly visible in Fig. \ref{fig1}. It shows 
\(\Delta\mathcal{E}\) for different models versus their magnetic Reynolds 
number. Apart from the aforementioned difference between the time-dependent 
and the stationary models, a dependence of \(\Delta\mathcal{E}\) on \(Rm\)
cannot be inferred.    

In addition, the growth rate of the leading eigenmode of \(\tens{D}\), 
\(\lambda=\Re(\sigma)\), may be taken as a measure for the accuracy of the 
mean-field description. Ideally, it is \(0\), whereas all overtones are 
highly diffusive \citep{schrinner10a,schrinner11b}. For numerical simulations, 
however, it is impossible to hit the critical point exactly. The growth rates 
of the fundamental mode for all models are given in Table \ref{table:1} in 
units of \(\eta/L^2\). Note that the turbulent diffusivity largely exceeds the 
molecular one. For model 2, for instance, we find values 
of \(\beta-\)components up to \(63\eta\). Thus, 
\(1/\lambda\approx1/7\, L^2/\eta\) for model 2 is in fact much larger than 
one effective decay time and the fundamental mode is already near its critical 
state. With decreasing \(\Delta\mathcal{E}\), the growth rates
become even smaller and the values closest to zero have been obtained 
for models 5, 6 and 7.  

\begin{figure}
  \centering
  \includegraphics{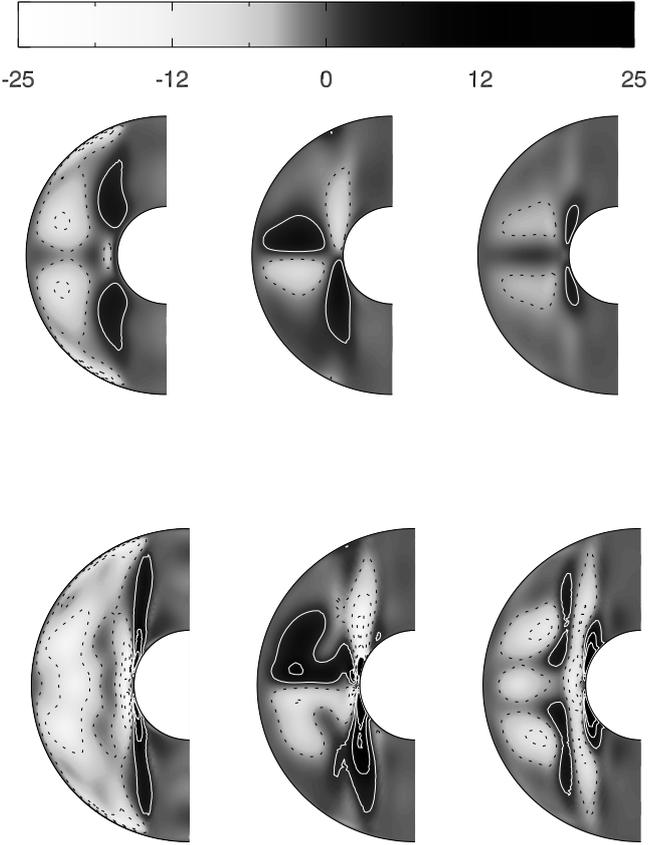}
  \caption{Contour plots of \(\mathcal{E}_r\), \(\mathcal{E}_\theta\), and 
  \(\mathcal{E}_\phi\) (top line, from left to right) and 
  \(\mathcal{E}^{(1)}_r\), \(\mathcal{E}^{(1)}_\theta\), and 
  \(\mathcal{E}^{(1)}_\phi\) (bottom line, from left to right) for model 7
  in units of \((\nu/L)\,(\varrho\mu\eta\Omega)^{1/2}\). In these units, 
  contour lines correspond to \(\pm 3,\pm 9,\pm 15\). 
  }
  \label{fig2}
\end{figure}

For model 7, the best parameterization was obtained and
\(\Delta\mathcal{E}=1.1\) was achieved. This still is clearly larger 
than for the simulation of rotating magnetoconvection for which 
\(\Delta\mathcal{E}=0.28\) was found by \cite{schrinner07}. Figure 
\ref{fig2} displays \(\vec{\mathcal{E}}\) in comparison to 
\(\vec{\mathcal{E}}^{(1)}\). Differences are visible in all components, but 
also principal similarities.  

\begin{figure}
  \centering
  \includegraphics{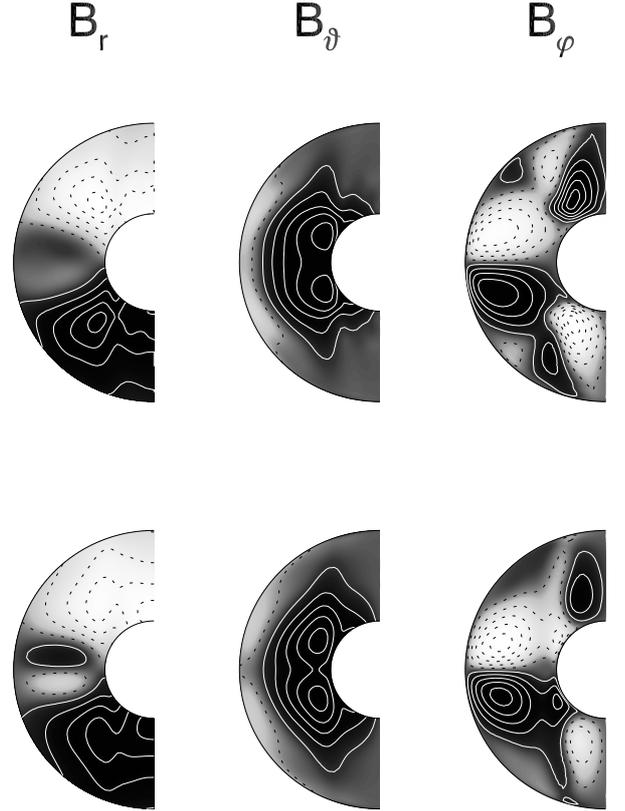}
  \caption{Comparison between direct numerical simulations and mean-field 
           calculations for model 7. Bottom line: azimuthally and 
           time-averaged magnetic field obtained by solving 
           (\ref{eq18}-\ref{eq22}). 
           Top line: fundamental eigenmode resulting from (\ref{eq32}). Each
           component has been normalised with respect to its absolute maximum.
           Therefore, the greyscale coding varies from \(-1\), white, to 
           \(+1\), black, and the contour lines correspond to 
           \(\pm 0.1,\,\pm 0.3,\,\pm 0.5\,\pm 0.7,\pm 0.9\).}
  \label{fig3}
\end{figure}

Figure \ref{fig3} compares the azimuthally and time-averaged magnetic field 
resulting from direct numerical simulations with the fundamental eigenmode 
of \(\tens{D}\) for model 7. Apart from small differences, direct numerical 
simulations and mean-field calculations agree very well for this example.


\section{Discussion}
We do not derive mean-field coefficients for time-averaged 
mean fields in a formal sense. Instead, the comparisons in Fig. \ref{fig2} 
and Fig. \ref{fig3} are motivated by assumptions i) and ii) in 
Section \ref{sec:6}. They are reasonably well fulfilled for the
simulations considered here. For model 7, \(\vec{v}\) time-averaged over 
30 magnetic diffusion times drops to \(0.1\%\) of \(\vec{V}\). A similar 
decrease has been found for \(\vec{b}\). Moreover, the axisymmetric flow
and the axisymmetric magnetic field show comparatively
little time variation. The time-dependent components of 
\(\overline{\vec{V}}\) and \(\overline{\vec{B}}\) are of the order of \(8\%\) 
\((37\%)\) and \(17\%\) \((23\%)\) of the total (axisymmetric) flow and 
the total (axisymmetric) magnetic field, respectively. In other words, 
time-averaging results in  axisymmetric fields and azimuthal averaging 
reduces somewhat the time variability. Combining both leads to an extended 
averaging and thus to an improved scale separation.  
\begin{figure}
  \centering
  \includegraphics{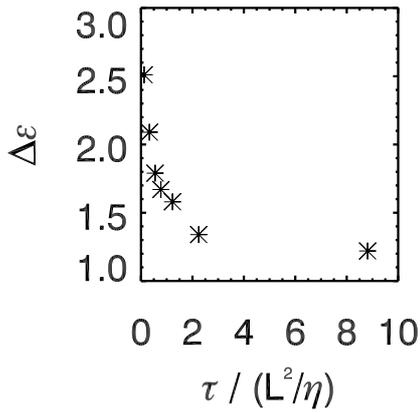}
  \caption{\(\Delta\mathcal{E}\) for model 5 varying with increasing 
  time-averaging period \(\tau\).}
  \label{fig8}
\end{figure} 
This finding is also supported by Fig. \ref{fig8} 
and Fig. \ref{fig9}. Figure \ref{fig8} shows a noticeable decrease 
of \(\Delta\mathcal{E}\) for model 5, if the time-averaging period 
\(\tau\) is increased. \(\Delta\mathcal{E}\) drops rapidly 
until \(\tau\approx 2 L^2/\eta\); an extension of the time-averaging 
interval over more than 2 magnetic diffusion times improves the scale 
separation only slightly. Moreover, figure \ref{fig9} compares the residual 
with the mean magnetic field for model 1 (upper panel) 
and model 7 (lower panel). In contrast to model 1, the time-dependent residual 
and the time-averaged mean magnetic field of model 7 vary on fairly 
different length scales. Thus, a better scale separation for 
the time-dependent models seems to be plausible.
\begin{figure}
  \centering
  \includegraphics{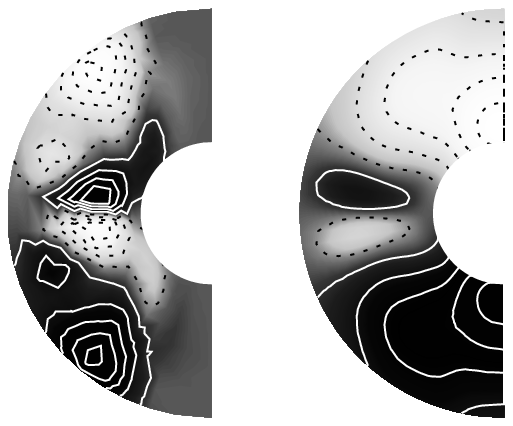}
  \includegraphics{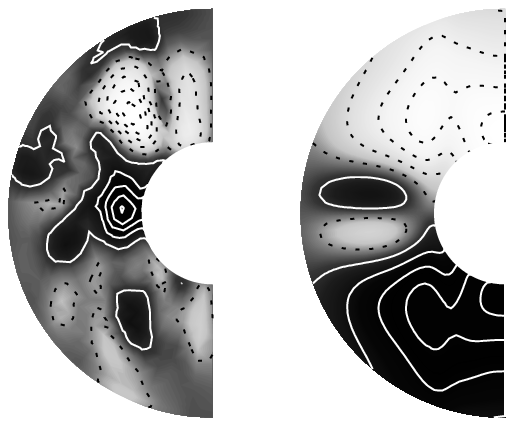}
  \caption{Upper panel: non-axisymmetric radial magnetic field at arbitrary 
   longitude (left) and azimuthally-averaged radial magnetic field (right) 
   for model 1. Lower panel: non-axisymmetric and time-dependent radial 
   magnetic field at arbitrary longitude and time (left) and azimuthally and 
   time-averaged radial magnetic field (right) for model 7. The style of the 
   contour plots is the same as in Fig. \ref{fig3}}
  \label{fig9}
\end{figure} 
Consequently, \(\Delta\mathcal{E}\) drops drastically from model 1 to model 2.
However, there are still noticeable differences between \(\vec{\mathcal{E}}\) 
and \(\vec{\mathcal{E}^{(1)}}\) in Fig. \ref{fig2} for model 7. 
From this comparison alone, it would be difficult to decide, whether the 
parameterization of \(\vec{\mathcal{E}}\) is already satisfactory.  
Further support for its reliability comes from the good agreement between 
direct numerical simulations and mean-field modeling in Fig. \ref{fig3}. 
Contour plots of all three components of the leading eigenmode of 
\(\tens{D}\) agree well with corresponding results from direct numerical 
simulations. Moreover, the growth rates of the leading eigenmodes are close 
to zero, as it has to be expected for stochastically stationary dynamos in 
the kinematically stable regime. 

Dynamo action in stars and planetary cores most probably takes place on
a wide range of length and time scales. Due to computational limitations, 
numerical simulations do not cover the whole range of scales possibly 
involved. Global dynamo simulations focus on large scales only, and small-scale
dynamo action as reported for box simulations 
\citep[e.g.][]{voegler07} is typically not present.
Therefore, the component of the electromotive force independent of the
mean magnetic field, i.e. \(\vec{\mathcal{E}}^{(0)}\) in (\ref{eq8}), is zero 
for most global dynamo simulations. The possibility of a contribution to 
\(\vec{\mathcal{E}}\) from small-scale dynamo action has been 
intensively discussed in the literature
\citep[e.g.][]{raedler76,raedler00,raedler07}. Recently, it has been used to 
argue against the applicability of mean-field theory \citep{cattaneo09}. 
However, conceptual difficulties which might result 
from simultaneous small- and large-scale dynamo action are at present not 
relevant for the approach followed here. 
        
The problem of scale separation has been addressed in this study with
regard to spatial scales, only. This seems to be appropriate because we intend 
to  describe stochastically stationary mean-fields with zero growth rate 
resulting from self-consistent dynamo simulations. More generally, time 
derivatives of the mean field in the expansion of \(\vec{\mathcal{E}}\) have 
to be taken into account. Similarly to non-local effects \citep{branden08}, 
also memory effects of the flow may influence the parameterization of 
the electromotive force \citep{hubbard09,hughes10}.   

Mean-field theory as presented here is intrinsically a kinematic approach. 
In general, an eigenvalue calculation based on (\ref{eq32}) will lead to 
growing modes, even if the mean-field coefficients are derived from a 
saturated velocity field \citep{tilgner08,cattaneo09b}. In such a situation, 
mean-field models might not match direct numerical simulations, unless a 
more self-consistent extension of the theoretical framework and the test-field 
method is applied \citep{cour10,RB10}. However, the class of chaotically 
time-dependent dynamos we considered is kinematically stable. 
\cite{schrinner10a} identified a regime of fast rotating dynamos which are 
dominated by only one dipolar mode at marginal stability, whereas all 
overtones are highly diffusive. A saturated velocity field from this class
of dynamos does not lead to exponential growth of the magnetic field in a 
corresponding kinematic calculation. Consequently, the mean-field approach
based on (\ref{eq32}) is applicable for this dynamo regime, as confirmed 
once more in this study. In particular for models beyond this regime, 
the reliability of the mean-field approach presented here is not guaranteed
and has to be made plausible by a comparison with direct numerical simulations 
\citep[e.g.][]{schrinner11a}.

\section{Summary}
The mean-field description of a (quasi-)stationary dynamo suffered from a poor
scale-separation and therefore from an insufficient parameterization of 
the electromotive force \citep{schrinner07}. For the chaotically 
time-dependent dynamos considered here, both improves a lot, if a combined 
azimuthal and time average is applied. The more accurate parameterization 
of \(\vec{\mathcal{E}}\) leads to a good agreement between mean-field models 
and direct numerical simulations: Field topologies and growth rates 
resulting from both approaches are very similar. 

In conclusion, the test-field method to
determine mean-field coefficients from direct numerical simulations proves 
to be reliable for chaotically time-dependent dynamos, too. Mean-field theory 
may serve as a powerful tool to analyse dynamo processes in global models 
resulting from direct numerical simulations.  

\begin{acknowledgements}
MS is grateful for financial support from the ANR Magnet project. 
The computations were carried out at the French national computing 
center CINES.
\end{acknowledgements}

\bibliographystyle{aa}
\bibliography{schrinner}

\end{document}